# Transition from a simple yield stress fluid to a thixotropic material

A. Ragouilliaux [1,2], G. Ovarlez [1], N. Shahidzadeh-Bonn [1], Benjamin Herzhaft [2], T. Palermo [2], P. Coussot [1]

[1] Université Paris-Est, Laboratoire Navier (UMR CNRS-ENPC-LCPC)
2 Allée Kepler 77420 Champs sur Marne, France
[2] IFP, 1-4 Av. du Bois Préau, 92852 Rueil Malmaison, France

**Abstract**: From MRI rheometry we show that a pure emulsion can be turned from a simple yield stress fluid to a thixotropic material by adding a small fraction of colloidal particles. The two fluids have the same behavior in the liquid regime but the loaded emulsion exhibits a critical shear rate below which no steady flows can be observed. For a stress below the yield stress, the pure emulsion abruptly stops flowing, whereas the viscosity of the loaded emulsion continuously increases in time, which leads to an apparent flow stoppage. This phenomenon can be very well represented by a model assuming a progressive increase of the number of droplet links via colloidal particles.

**PACS**: 82.70.-y, 83.60.Pq, 61.43.Hv

## I. Introduction

Various systems such as foams, emulsions or colloidal suspensions, are made of elements interacting via soft interactions, i.e. which vary progressively with the distance between the elements [1-2]. Their jammed character, namely their ability to undergo a solid-liquid transition when submitted to a sufficient stress, arouses the interest of physicists as it appears typical of a fourth state of matter with some analogy with glass behavior [3]. Their rheological behavior is usually described with the help of simple yield stress models, which may yield velocity profiles with unsheared regions when the stress distribution is heterogeneous. Recently some works showed that the flow of such materials exhibits "true shear-banding", i.e. two regions with significantly different shear rates in geometries in which the shear stress is homogeneous [4]. Such a behavior is not predicted by usual yield stress models for which the transition between the solid and the liquid phases is smooth.

Two questions thus emerge: (i) how can a material become a "shear-banding" material? and (ii) how does the transition towards shear-banding occur in time for a given material? The first question was recently addressed by Bécu et al. [5] who showed that a simple yielding emulsion can exhibit shear-banding if it is made "adhesive", i.e. the droplets develop some adhesive interactions. Here we focus on question (ii), namely the time-dependency of this shear-banding effect and its origin, as it is likely that the transition from the purely solid (starting from rest) or purely liquid (starting from an intense preshear) regimes towards shear-banding is a complex phenomenon from which useful information for our understanding of these materials could be extracted. Remark that there is certainly some interplay of these phenomena with the thixotropic character of these fluids (time variations of the apparent viscosity). For example it was shown that a clay-water system can exhibit very different behavior types, and in particular shear-banding or even fractures, depending on the time it spent at rest before undergoing a stress [6].

Here we compare the behavior of an emulsion either pure or loaded with colloidal particles. In order to have a straightforward appreciation of the effective behavior of these materials we observe their flow characteristics in time with the help of MR (Magnetic Resonance) velocimetry. We show that the pure emulsion can be considered as a simple yield stress fluid with negligible thixotropy while the loaded emulsion is significantly thixotropic. Both fluids have the same behavior in the liquid regime and the thixotropy of the loaded emulsion is due to a progressive aggregation of droplets via colloidal particle links, which tends to increase the viscosity of the material which finally stops. As a result the apparent yield stress (below which no steady flow can be obtained) of the loaded emulsion is larger than the pure emulsion. We finally show that during stoppage the shear rate exhibits a specific evolution which is consistent with the predictions of a model assuming a progressive aggregation of droplets in time.

## II. Materials and procedures

We prepared the pure emulsion by progressively adding the water in an oil-surfactant solution (Sorbitan monooleate, 2%) under high shear. The size distribution of water droplets was almost uniform around a diameter of $1~\mu m \pm 20\%$. The water droplet concentration was 70%, a concentration larger than the maximum packing fraction of uniform spheres, which means that the droplets were probably slightly deformed. However the droplet deformation due to flow can be neglected during all the flows considered here since, due to the very small droplet size, the interfacial stress is much larger than the stress due to flow. As a consequence the average shape of droplets can be considered as constant during all our tests. The loaded emulsion was prepared by mixing the surfactant and the colloidal particles (slightly flexible, large aspect ratio clay particles (*Bentone 38*, Elementis Specialties company) with a mean diameter: $\approx 1\mu$m and a thickness $\approx 0.01\mu$m) at a solid volume fraction of 3% in the oil, then progressively adding the water in this suspension under high shear. Remark that the clay-oil suspension (with the same solid volume fraction) alone is a simple liquid (which shows no yield stress) of low viscosity ($5.10^{-3}$ Pa.s) whereas the clay-emulsion mixture is a much more viscous pasty material even at low water fractions (e.g. at a volume fraction of 10% its apparent viscosity is of the order of $5.10^{-1}$ Pa.s). As a consequence it is likely that the clay particles tend to form links between neighbouring droplets, which thus form aggregates or even a continuous network throughout the material in some cases. The exact nature of these links is not yet understood.

We carried out experiments with a wide-gap Couette rheometer (inner cylinder radius: 4cm; outer cylinder radius: 6cm; height of sheared fluid: 11cm) inserted in a MRI set-up. We imposed the rotation velocity of the inner cylinder and measured the torque and the velocity profile ($V(r)$ in which $r$ is the distance from the central axis) inside the sample at successive times by magnetic resonance velocimetry. For a given rotation velocity the torque ($M$) was constant in time even if for the loaded emulsion the velocity profile significantly evolved (see below). The principle of the magnetic resonance velocimetry technique used was described elsewhere [7]. The typical time needed to get one velocity profile was 15s. The uncertainty on the local velocity measurements was typically $5.10^{-5}$ m/s. Slight wall slip effects possibly occurred but did not affect the transient measurements as they remained constant for a given torque and did not affect our rheological interpretations based on measurements inside the sample.

We also carried out NMR density tests with the help of a new technique [8] for separating the oil and the water signals, which make it possible to detect within 1% some heterogeneity of the water fraction resulting from migration effects. No migration was observed in the pure emulsion. With the clayey emulsion no such migration effect was observed during the first 20min. of flow but a significant migration developed beyond 30min. of flow. In the following we present only the data corresponding to the first stage (with negligible migration). We also carried out conventional rheometrical tests (dynamic tests and creep tests) with a *Malvern* rheometer equipped with a thin-gap Couette geometry (inner radius: $r_1 = 17.5$mm; outer cylinder radius: $r_2 = 18.5$mm; height: $h = 45$mm) with rough surfaces (roughness: 0.18mm).

## III. Experimental results

A. Solid regime

From creep tests we find that both materials exhibit a solid and a liquid regimes: at low stress the induced deformation tends to saturates (solid regime), whereas at high stress they tend to flow at a constant rate (liquid regime). This is the typical behavior observed for pasty materials [9]. Let us first compare the material properties in their solid regime from dynamic tests (frequency: 1Hz, strain amplitude: 1%). For the pure emulsion the elastic modulus ($G'$) slightly increases with time (see inset of Fig.1) (i.e. of 7% after 3000s), which means that, as a first approximate, it negligibly ages in its solid regime. Thus the possible evolution of the droplet configuration in time only induces a slight increase of the network strength. In contrast for the loaded emulsion $G'$ significantly increases in time (i.e. of 70% after 3000s). Thus, as time goes on, more and more clay links form between droplets, which strengthens the network of droplet contacts.

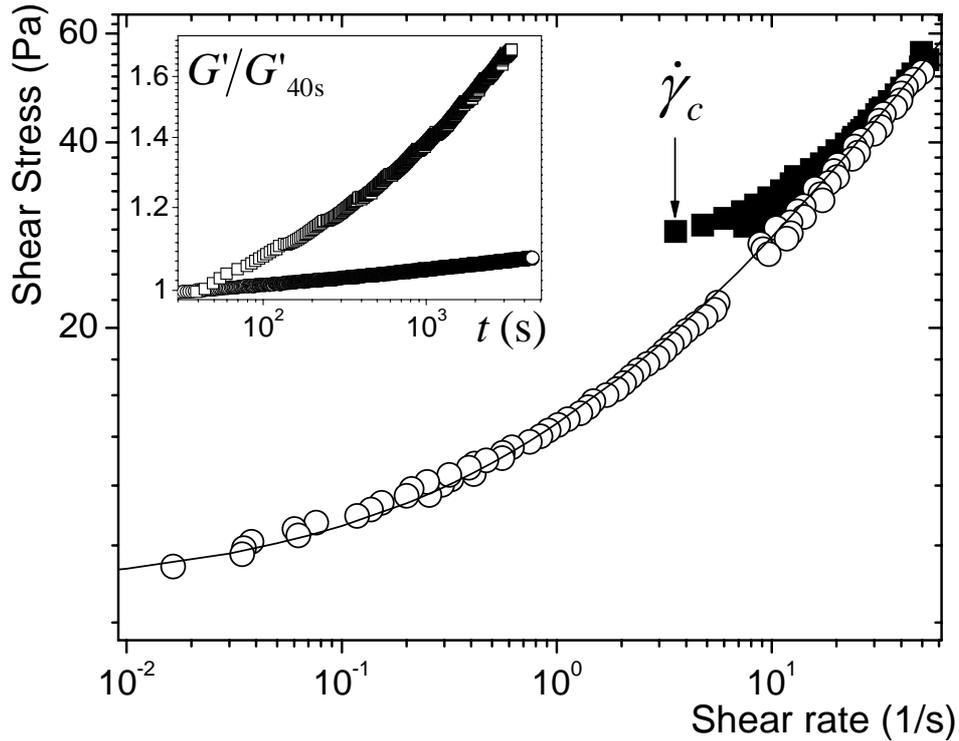

Figure 1: Steady-state shear stress vs shear rate for the pure (circles) and clayey (squares) emulsions as deduced from local measurements within the wide-gap rheometer under different torque values. The solid line is a Herschel-Bulkley model ($\tau = \tau_e + K\dot{\gamma}^n$) fitted to data (with $\tau_e = 7.5\text{Pa}$, $K = 6.5\text{Pa.s}^{0.5}$ and $n = 0.5$) for the pure emulsion. Inset: Elastic modulus (scaled by the value reached at 40s) of the pure (circles) and clayey (squares) emulsions as a function of time.

B. Liquid regime

Let us now consider the behavior in the liquid regime. In that aim we preshear the material at a large velocity and leave it at rest few seconds, then we impose a given rotation velocity and observe the velocity profile inside the gap. During a short initial period (say less than few seconds) the rotation velocity varies as a result of inertia and solid-liquid transition effects. Afterwards, for the pure emulsion the velocity profile is constant. This means that this material exhibits negligible thixotropy in the liquid regime: the specific (average) droplet configuration associated with a given flow rate is reached almost immediately. In contrast, with the loaded emulsion the velocity profile significantly evolves in time: it progressively moves towards the central axis (cf. Fig.2) and finally reaches a steady shape. During this process the outer region apparently progressively stops flowing but there still might remain some slow flow at velocities smaller than the uncertainty on these measurements.

In a Couette geometry the shear stress distribution ($\tau$ as a function of $r$) in the material is well known from the momentum equation and we have $\tau(r) = M/2\pi h r^2$, i.e. at a given torque level the stress decreases with the distance from the axis. Our observations concerning the velocity profiles under a constant torque imply that in the regions of low stresses (near the outer cylinder) the material tends to stop flowing while in the regions of large stresses (near the inner cylinder) it tends to reach a steady-state flow. This effect is reminiscent of the viscosity bifurcation effect observed with various material types [10]. However here, in contrast with previous studies, we have a straightforward information concerning the effective (local) shear rate in time. Indeed we can compute the local shear rate ($\dot{\gamma}$) for any stress (associated with a given distance $r$) at any time: we have $\dot{\gamma} = r\, \partial(V/r)/\partial r$. Looking now at the evolutions in time of the shear rate under different stress values (i.e. at different distances) we effectively observe the viscosity bifurcation effect (see Fig.3): for $\tau > \tau_c$ the shear rate tends to a constant value, for $\tau < \tau_c$ the shear rate progressively tends to zero, with a characteristic time increasing with $\tau$. As a consequence there exists a critical shear rate ($\dot{\gamma}_c$), associated with $\tau_c$, below which no steady-state flows can be obtained, the material tends to stop. From our data we get $\dot{\gamma}_c \approx 3 \text{s}^{-1}$.

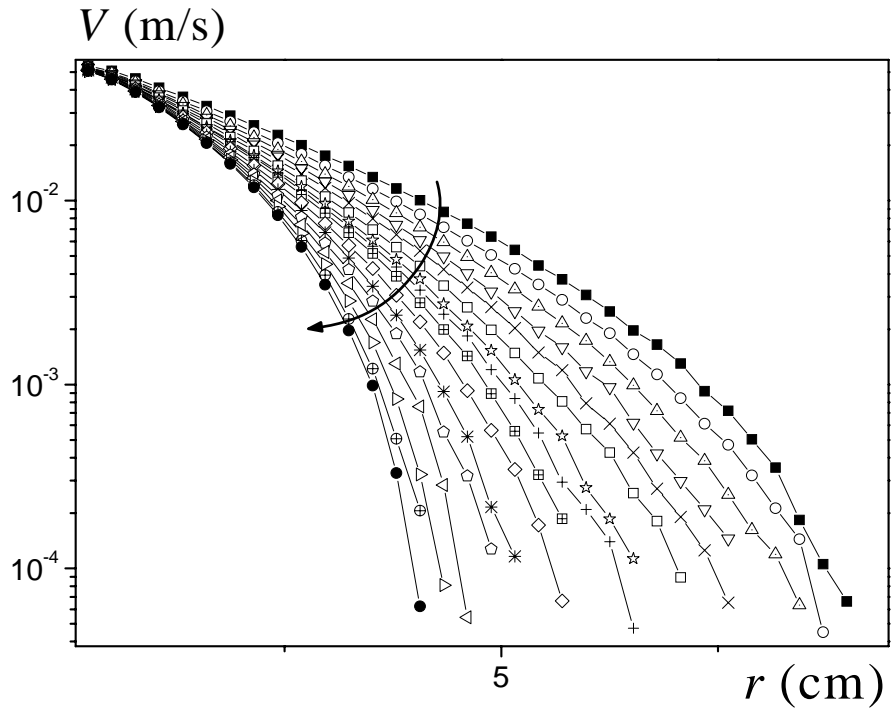

Figure 2: Velocity profiles at different times in the wide-gap Couette geometry as measured by MR velocimetry for the clayey emulsion for a rotation velocity of 15rpm: (from right to left) 30, 60, 90, 120, 150, 180, 210, 240, 270, 330, 390, 450, 540, 660, 780, 900s. The velocity values below the uncertainty ($5.10^{-5}$ m.s$^{-1}$) are not represented, which explains why there is no data beyond some critical distance for each profile.

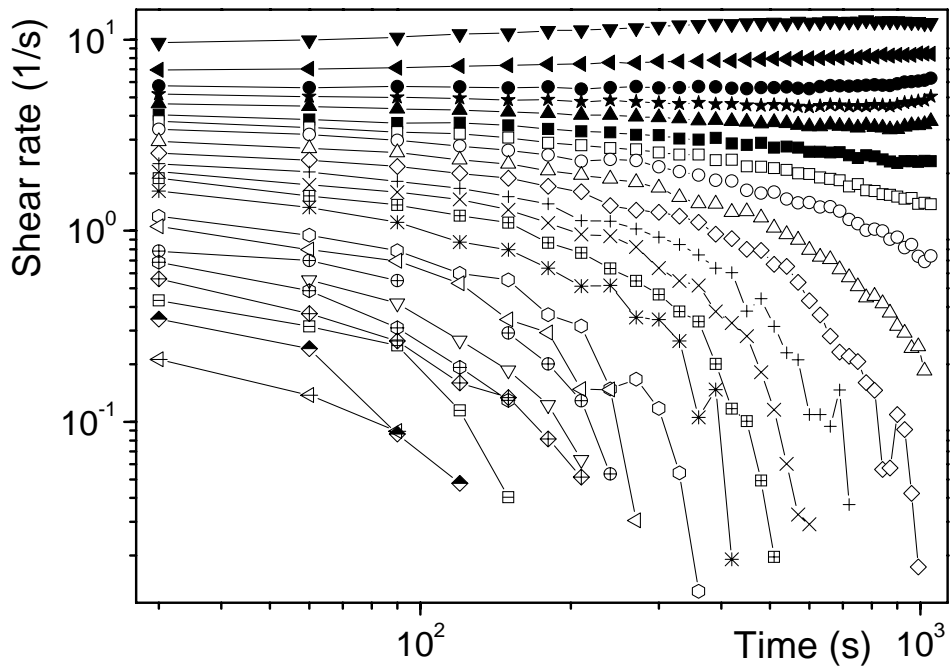

Figure 3: Effective shear rate as a function of time for different local stresses as determined from the MR velocity profiles for the clayey emulsion for a rotation velocity of 15rpm: (from top to bottom) 34.2, 31.7, 30.1, 29.4, 28.7, 28, 27.3, 26.7, 26.1, 25.5, 24.9, 24.3, 23.8, 22.8, 22.3, 21.8, 21.4, 20.9, 20.5, 20.1, 19.7, 18.9 Pa.

## IV Discussion

A. Data analysis

Let us now compare the constitutive equation of these materials in the liquid regime in steady state as it can be deduced from a set of experiments under different torques from MRI data, by computing the local shear rate and shear stress for each value of $r$ as above described. The flow curve (steady state $\tau$ vs $\dot{\gamma}$) of the pure emulsion is that of a simple yield stress fluid, i.e. the shear stress tends to a plateau at low shear rates below which no steady flows are obtained and a Herschel-Bulkley model very well represents the flow curve (see Fig.1). The loaded emulsion exhibits a behavior similar to that of the pure emulsion for shear rates larger than say $10\text{s}^{-1}$. This means that the clay links no longer play any role and suggest that there could be all broken.

The striking difference between the two materials concerns the minimum stress level allowing steady state flow: the yield stress of the clayey emulsion is significantly larger than that for the pure emulsion and just beyond this yield stress the fluid flows at a shear rate larger than $\dot{\gamma}_c$. Forgetting the short plateau of the flow curve for the clayey emulsion we get the following rough picture: both materials have a similar behavior in the liquid regime but the flow curve of the clayey emulsion is truncated at a critical shear stress larger than the yield stress of the pure emulsion, and no stable flows can be obtained at a shear rate below the critical value associated with this yield stress. For such materials shear-banding will occur in shear flows at apparent shear rates smaller than $\dot{\gamma}_c$. This behavior is reminiscent of the trend observed by Becu et al. [5] with an adhesive emulsion, but here we get a precise picture of the different fluid behavior types at a local scale and we show that the difference finds its origin in some time-dependent effect.

With regards to our scheme of the microstructure it is natural to consider that this behavior is the result of a dynamic effect: a progressive transition from flow to stoppage, which is likely due to the progressive formation of larger aggregates. Thus as a first approximation we can consider that all the clay links are broken during a rapid shear, so that the clay fraction plays a negligible role (in agreement with our observation of the similar behavior of the two materials in the liquid regime), but below a critical stress the number of links increases significantly, which slows down the flow so that the fraction of links further increases and so on until full stoppage.

Previous studies of the thixotropic properties in the liquid regime generally relied on macroscopic data under poorly controlled conditions. Here, from MR velocimetry we have straightforward information concerning the local viscosity variations in time (see Fig.3). We can remark that the shear rate vs time curves for different stress values below $\tau_c$ look similar in a logarithmic scale: they start with a plateau at short time then rapidly decrease after some time, which decreases as the stress decreases. This similarity definitely appears when plotting the shear rate vs the time scaled by appropriate factors, respectively $\dot{\gamma}_0(\tau)$ and $\theta(\tau)$ (see Fig.4). This is reminiscent of an effect recently observed [11]: for various suspensions the elastic modulus vs time curves after the liquid-solid transition under different stress fall along a master curve when the elastic modulus and the time are scaled by appropriate functions of the stress. Here we get the following general result: for a stress imposed below the critical stress the shear rate evolves as $\dot{\gamma} = \dot{\gamma}_0 f(t/\theta)$, with $\dot{\gamma}_0 \approx [(\tau - \tau_0)/(\tau_c - \tau_0)]\dot{\gamma}_c$ and $1/\theta \approx k(\tau_c - \tau)$. This similarity of the behavior strongly suggests that the structural state of the material is a function of $\xi = t/\theta$ only, which thus can be considered as describing the physical age, i.e. the state of structure, of the material during such a test. It is remarkable that the evolutions of this parameter only depend on the stress. This contrasts with usual models (e.g. [6]) in that field which assume that the structure parameter essentially depends on the shear rate.

From the above equations we deduce the apparent behavior of the material for all $\xi$:

$$\tau = \tau_0 + (\eta_0/f(\xi))\dot{\gamma}(\xi) \qquad (1)$$

where $\eta_0 = (\tau_c - \tau_0)/\dot{\gamma}_c$. In this equation the first term, i.e. $\tau_0$, is independent of the structural state and thus can be considered as reflecting the yield stress of the clayey emulsion when there is no droplet-clay links. This yield stress takes its origin in the jammed droplets network with clay possibly affecting the interfacial tension between oil and water, which explains that it may differ from the yield stress of the pure emulsion. The second term (the "viscous term") is associated with both the viscous effects and the droplet-clay links in the emulsion. It is worth noting that the behavior described by equation (1) does not correspond to a Bingham model. Indeed the factor of the shear rate $(\eta_0/f(\xi))$, i.e. the so-called "plastic viscosity" which is constant in the Bingham model, in our model depends on the structural state $\xi$ and thus depends on the shear stress.

**B. Model**

The above analysis suggests that one may see the clayey emulsion as a suspension of linked elements in an interstitial homogeneous yield stress fluid. We will make two simplifying assumptions: (i) the stress in this suspension may be expressed as the sum of a "yielding" term strictly due to the jamming-unjamming process during flow and a simple viscous term with an apparent viscosity ($\eta$), resulting from the additional viscous dissipation due to flow and function of the actual concentration of aggregates; (ii) the presence of aggregates negligibly affects the yielding part. Such a picture is qualitatively in perfect agreement with the conclusions of our data interpretation (see above). In this context the viscous term may be found from the viscosity of a Newtonian suspension in a liquid of viscosity $\eta_0$.

We assume that at a given time the droplet-clay links form more or less large (rigid) networks which trap some volume of the interstitial liquid. Various models exist for the viscosity of rigid spherical inclusions [12] but it seems more relevant to use the expression found from energy dissipation considerations in the context of purely viscous suspensions of aggregative particles [13]: $\eta = \eta_0 (1 - \phi_{agg})\left(1 - (\phi_{agg}/\phi_m)\right)^{-2}$. Here $\phi_{agg}$ is a parameter which represents the volume fraction of rigid inclusions formed by interstitial liquid trapped in clay-droplet networks. The specific maximum packing fraction $\phi_m$ used in the Mills model [13] is equal to 0.57. A different choice for this value would slightly change the value of the fitting parameter in the following analysis but not the qualitative results. This viscous term can be associated with $\eta_0/f(\xi)$ since both terms describe the apparent viscosity of the suspension in a simple liquid of viscosity $\eta_0$, which provides a correspondence between $\phi_{agg}$ and $\xi$. For example a kinetics relation of the form $\phi_{agg} \propto \xi^{3/5}$ makes it possible to get a good fit of the above equation to the master curve (see Fig.4). Remark that the aggregate size tends towards infinity at stoppage, so that the above theory no longer holds and the corresponding stress contributes to increase the yield stress of the material, which explains that the apparent yield stress of the clayey emulsion is larger than pure emulsion.

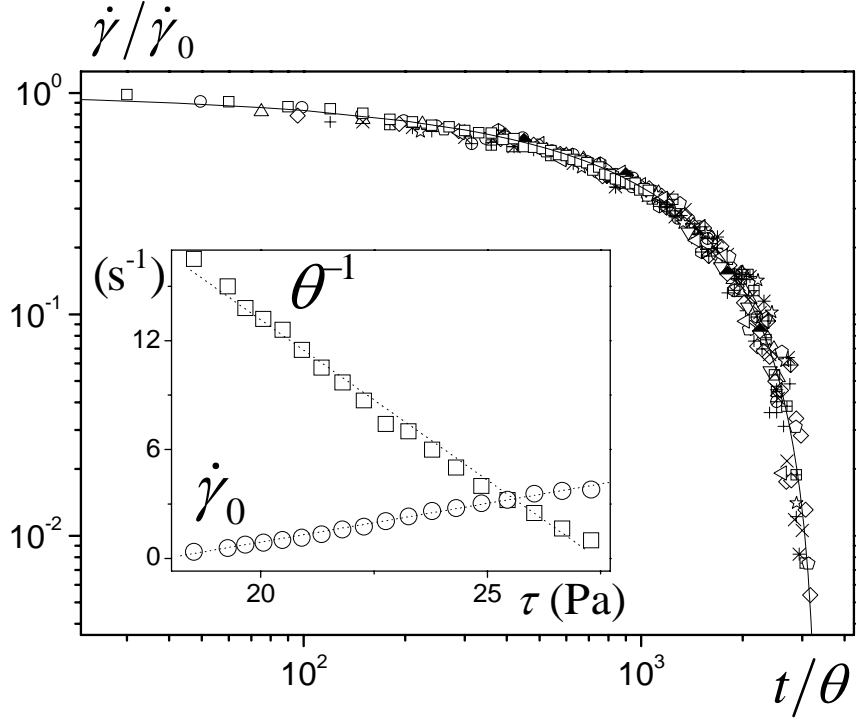

Figure 4: Shear rate (data of Figure 3 for $\tau \leq 27.3\,\text{Pa}$) scaled by a factor $\dot{\gamma}_0(\tau)$ as a function of the time scaled by a factor $\theta(\tau)$ for the clayey emulsion. The continuous line is the Mills model fitted to data (with $\phi_{agg} = 0.00432\xi^{3/5}$). The inset shows the values of these two parameters as a function of the stress. The dotted lines are the empirical models: $\dot{\gamma}_0 \approx [(\tau-\tau_0)/(\tau_c-\tau_0)]\dot{\gamma}_c$ and $\theta^{-1} \approx k(\tau_c-\tau)$ with $\tau_0 = 17.9\,\text{Pa}$, $k = 1.75\,\text{s.Pa}^{-1}$, and $\tau_c = 27.5\,\text{Pa}$.

The above equation provides the relation between the effective advancement of the aggregation (via $\phi_{agg}$) and the time and the stress applied. It is remarkable that the stress is solely involved in this relationship whereas in general for such physical phenomena the characteristic time is proportional to $1/\dot{\gamma}$. One finds a possible beginning of explanation of such an effect in the usual description of simple liquids [14], glassy [15] or jammed systems [16] in which the elements are assumed to be in a potential well with an energy barrier of depth $\Delta E$. We can assume that such a frame applies for the clay particles involved in clay-droplet links. The probability of escape from this potential well due to thermal agitation is $\exp-\Delta E/k_B T$ per attempt, in which $k_B$ is the Boltzmann constant and $T$ the temperature. In this context the influence of stress is to decrease the energy barrier by a factor $\tau d$ in which $d$ is a characteristic distance of the system, and thus to foster the fluid flow. In order to have the possibility to form a clay bridge two droplets must be neighbours. Over a time of flow $\Delta t$ the number of occurrence of droplet neighbouring is proportional to $\dot{\gamma}$ but the time during which two droplets can be considered as neighbours is inversely proportional to $\dot{\gamma}$, so that the rate of flow has no influence on the number of links (this result is valid only because we neglected possible link breakages). To sum up we find that the rate of variation of the number of links increases with the stress and is independent of the shear rate, in agreement with our experimental results. However a quantitative prediction of the effect of stress on $\theta$ requires a more complete theoretical approach.

## V. Conclusion

We have shown that the transition from a non-thixotropic to a thixotropic emulsion occurs by the disappearance of steady flows below a critical stress associated with a critical shear rate. Here this process is well explained by the increase of droplet links via clay particles. These effects might be general for jammed systems for which the aging is due to reversible particle aggregation.

## References

[1] P. Coussot, *Rheometry of pastes, suspensions and granular materials* (Wiley, New York, 2005)

[2] V. Trappe et al., *Nature*, 411, 772 (2001)

[3] A.J. Liu, and S.R. Nagel, *Nature*, 396, 21 (1998)

[4] F. Pignon, A. Magnin, and J.M. Piau, *J. Rheol.*, 40, 573-587 (1996); P. Coussot et al. *Phys. Rev. Lett.*, 88, 218301 (2002)

[5] L. Bécu, S. Manneville, and A. Colin, *Phys. Rev. Lett.*, 96, 138302 (2006)

[6] P. Coussot et al., *Phys. Fluids*, 17, 011704 (2005)

[7] J.S. Raynaud et al., *J. Rheol.*, 46, 709-732 (2002)

[8] G. Ovarlez, S. Rodts, P. Coussot, J. Goyon, A. Colin, submitted to *J. Colloid and Interface Science* (2007)

[9] P. Coussot et al., *J. Rheol.*, 50, 975-994 (2006)

[10] F. Da Cruz et al., *Phys. Rev. E*, 66, 051305 (2002)

[11] G. Ovarlez, and P. Coussot, *Phys. Rev. E,* 76, 011406 (2007).

[12] I.M. Krieger, and T.J. Dougherty, *Trans. Soc. Rheol.*, 3, 137-152 (1959); D. Quemada, *J. Mécan. Théor. Appl.*, Special Issue, 267-288 (1986)

[13] P. Mills *J. Phys. III.,* 6, 1811 (1996)

[14] S. Glasstone, K. J. Laidler, and H. Eyring, *The Theory of Rate Processes* (McGraw-Hill, New York, 1941)

[15] C. Monthus and J.P. Bouchaud, *J. Phys. A: Math. Gen.*, 29, 3847 (1996); F. Varnik et al., *Phys. Rev. Lett.*, 90, 095702 (2003)

[16] P. Sollich et al., *Phys. Rev. Lett.*, 78, 2020 (1997); P. Hébraud and F. Lequeux, *Phys. Rev. Lett.*, 81, 2934 (1998); P. Sollich, *Phys. Rev. E*, 58, 738 (1998); C. Derec, A. Ajdari, and F. Lequeux, *Eur. Phys. J. E,* 4, 355 (2001); Y. L. Chen and K. S. Schweizer, *J.Chem.Phys.,* 120, 7212 (2004)